\documentclass[prl,twocolumn,superscriptaddress,showpacs,nofootinbib,preprintnumbers,amsmath,amssymb]{revtex4}

\usepackage[dvipdfm]{graphicx}
\usepackage{amssymb}
\usepackage{amsmath}
\usepackage{parskip}
\usepackage{epsfig}
\usepackage{dcolumn}
\usepackage{rotating}

\begin{document}

\preprint{IGPG-07/2-1}

\title{A fine tuning free resolution of the cosmological constant problem}

\author{Stephon Alexander}
\email{stephon@slac.stanford.edu}
\author{Deepak Vaid}
\email{deepak@phys.psu.edu}
\affiliation{Department of Physics,\\
Institute for Gravitational Physics and Geometry,\\
The Pennsylvania State University,\\
104 Davey Lab, University Park, PA,16802, U.S.A \\ }
\date{\today}

\begin{abstract}
In a recent paper we discovered that a fermionic condensate is
formed from gravitational interactions due to the covariant coupling
of fermions in the presence of a torsion-fermion contact interaction.
The condensate gap gives a negative contribution to the bare
cosmological constant.  In this letter, we show that the
cosmological constant problem can be solved without fine tuning of
the bare cosmological constant.  We demonstrate how a universe with
a large initial cosmological constant undergoes inflation, during
which time the energy gap grows as the volume of the universe.
Eventually the gap becomes large enough to cancel out the bare
cosmological term, inflation ends and we end up in a universe with
an almost vanishing cosmological term.  We provide a detailed
numerical analysis of the system of equations governing the self
regulating relaxation of the cosmological constant.
\end{abstract}


\maketitle

\section{Introduction}

There are many faces to the cosmological constant/dark energy
problem.  First, the naive perturbative theoretical evaluation of
the vacuum energy of all particles in the standard model gives a
result that disagrees with observations by 120 orders of magnitude
\cite{Carroll-2001}. Second, a confluence of cosmological and
astrophysical observations, such as the WMAP satellite
\cite{spergel-2006} and Type Ia supernovae \cite{astier-2006},
indicate that the cosmological constant or something very similar to
it, currently dominates the universe.

Perhaps the most striking aspect of the cosmological constant
problem is seen in the details of the inflationary
paradigm\cite{Brandenberger:2002wm}. Inflation is driven by a
constant part of the Energy-Momentum tensor of a scalar field, which
is indistinguishable from a pure cosmological constant.  Therefore,
any mechanism which relaxes the cosmological constant would also
prevent inflation from happening. One way out of this possible
conundrum is to do away with fundamental scalar fields, allow
inflation to occur with a large cosmological constant and
investigate any self consistent mechanism which negates the
cosmological constant to almost zero.  Such a mechanism would solve
all three cosmological constant problems:

\begin{itemize}

\item The cosmological constant would be dynamically relaxed due to the non-trivial dynamics of
inflation itself; it would be self regulatory.

\item Dark energy and the coincidence problem would be explained if a residual amount of
cosmological constant would be left over by the end of inflation.

\item Since inflation is not driven by a fundamental scalar fields, fine tuning of the cosmological
constant is no longer needed.

\end{itemize}

Attempts at tackling the this problem via cosmological condensates
include \cite{brandenberger-1997,brout-2003, arkanihamed-2004,
alexander-2004, alexander-2005}. More recently, Prokopec proposed a
mechanism involving a Yukawa coupling between a scalar field and
fermions \cite{prokopec-2006}.

A simple way to obtain inflation in the absence of matter is due to
the presence of a non-zero, positive cosmological term on the right
hand side of Einstein's equation:

\begin{align}
    G_{ab} = 8\pi G \Lambda_0 g_{ab}\nonumber \\
   \Rightarrow a(t) = a_0 (t) e^{\sqrt{\frac{\Lambda_0}{3}}t}
\end{align}

where we have used the FRW metric ansatz to obtain our solution.
$a(t)$ is the scale factor.  From the solution it is clear that the
Hubble rate $H = \sqrt{\frac{\Lambda_0}{3}}$.

While this simple model gives us an inflating universe it is clearly
not in line with reality because it does not predict an end to
inflation.  A way to get around this obstacle is to introduce
matter, traditionally scalar fields, into the picture. Then the
first Friedmann equation becomes:

\begin{equation}
    3\left( \frac{\dot a}{a} \right)^2 = \Lambda_0 + \frac{1}{2}
    \dot \phi^2 + V(\phi)
\end{equation}

where $\phi$ is the scalar field.

We also have the E.O.M for the scalar field:

\begin{equation}
    \ddot \phi + 3 \frac{\dot a}{a} \dot \phi + \frac{dV}{d\phi} = 0
\end{equation}

where $V(\phi)$ is the scalar field potential. Such models typically
require special initial conditions for the scalar field called the
"slow-roll" conditions. The scalar field must start off at a large
initial value and then start rolling slowly down an almost flat
potential. This results in an inflationary universe. After a
sufficient number of e-foldings, the scalar fields reaches the
steeper part of the potential where it decays via parametric
resonance leading to reheating and particle production after
inflation has ended.

Unfortunately, such models have several shortcomings:
\begin{itemize}
  \item The shape of the potential is arbitrary and we have no physical way of choosing
  the one that would correspond to our universe from an almost
  infinitely large family.
  \item We require the scalar field to start off at a large initial
  value. What mechanism would cause the scalar field to be "pumped
  up" to this value initially?
  \item The mass of the scalar field is an arbitrary parameter. It
  can be fixed once we fix the potential, but it remains a source of
  vagueness.
  \item Most importantly, from whence did this scalar field come
  . Perhaps if one tries to tackle this fundamental question head
  on the others might also be amenable to a solution.
\end{itemize}

In this letter we propose a dynamical solution to the CC problem
assuming only the Standard Model and General Relativity.  There are
no fundamental scalar fields to tune.  Therefore the universe will
be dominated by a large cosmological constant, which naturally
generates inflation.  The non-trivial observation here is that the
dynamics of inflation itself holds the key to relaxing the
cosmological constant without fine tuning. How is this possible? The
exponential time dependent behavior of de Sitter space
counterintuitively enhances correlations between fermion pairs.
These correlations become so strong that these fermions form a
Cooper pair.

In a recent paper \cite{alexander-2006}, we showed how the presence
of torsion and fermionic matter in gravity naturally leads to the
formation of a fermionic condensate with a gap which depends on the
3-volume.  In this letter we will analyze explicitly the dynamics of
the universe with a cosmological constant in the presence of this
gap. Numerical calculations then show that with a large initial
cosmological term and generic initial conditions for the scalar
field and its momenta, we obtain a universe which undergoes an
inflationary phase during which the gap grows as a function of
$a^3$, causing the effective cosmological term to diminish to a
small positive value.

In Section 2 we discuss the E.O.M for our system. In Section 3 we
present the numerical results and finally we conclude with some
discussion or our results and what they imply for our understanding
of inflation and the cosmological term.

\section{Friedmann and scalar equations}

We briefly summarize the steps that were taken in
\cite{alexander-2006}. We started with the Holst action for gravity
with fermions:

\begin{widetext}
\begin{equation}
    S_{H+D} = \frac{1}{2\kappa}\int d^{4}x\,e(\,e^{\mu}_{I}e^{\nu}_{J}R^{IJ}_{\mu\nu} - \frac{2}{3}\Lambda_0) -
    \frac{1}{2\kappa\gamma}\int d^{4}x\,e\,e^{\mu}_{I}e^{\nu}_{J}\star R^{IJ}_{\mu\nu} + \frac{i}{2}\int d^{4}x\,e\,(\bar{\Psi}\gamma^{I}e^{\mu}_{I}D_{\mu}\Psi -
    \overline{D_{\mu}\Psi}\gamma^{I}e^{\mu}_{I}\Psi)
\end{equation}
\end{widetext}

$e^\mu_I$ is the tetrad field. $R_{\mu\nu}^{IJ}$ is the curvature
tensor. The second term in the above equation is analogous to the
$\Theta$ term in Yang-Mills theory and is required if we want to
work with arbitrary values of the Immirzi parameter ($\gamma$).
After varying the action w.r.t the connection $A^\mu_{IJ}$ and
solving the Gauss constraint we which that $A^\mu_{IJ} =
\omega^\mu_{IJ} + C^\mu_{IJ}$\footnote{Which implies that the
torsion is non-zero}, where $\omega^\mu_{IJ}$ is the tetrad
compatible connection and $C^\mu_{IJ}$ can be expressed in terms of
the axial vector current:

\begin{equation}
    C_\mu^{IJ} = \frac{\kappa}{4}\frac{\gamma^2}{\gamma^2 + 1}j^M_a\left \{ \epsilon_{MK}{}^{IJ}e^K_\mu
            - \frac{1}{2\gamma}\delta^{[J}_M e^{I]}_\mu \right\}
\end{equation}

where $ j^M_a = \bar\Psi\gamma_5\gamma^M\Psi$. Inserting the torsion
into the first order action we find the resulting second order
action which now contains a four-fermi interaction and the tetrad is
the only independent variable, the connection having already been
solved for in the previous step.

\begin{equation}
    S[e,\Psi] = S_{H+D}[\omega(e)] -\frac{3}{2}\pi G \frac{\gamma^{2}}{\gamma^{2}+1} \int d^{4}x\,e (j_a^I)^2
\end{equation}

Then we did the 3+1 decomposition of the action to find the
Hamiltonian, which after making the ansatz of a FRW metric becomes:

\begin{eqnarray}
\label{ham}
    {\cal H} &=& -\frac{3}{\kappa}a^3H^2 +
    a^3\Lambda_0 + \frac{i}{a}\big(\psi_L^\dag\sigma^a\partial_a\psi_L - \psi_R^\dag\sigma^a\partial_a\psi_R\big) \nonumber \\
    &+& \frac{3\kappa}{32a^3}\frac{\gamma^2}{\gamma^2+1}\left[\psi_L^\dag\psi_L - \psi_R^\dag\psi_R\right]^2 = 0
\end{eqnarray}

We see that the right hand side is the sum of the gravitational,
Dirac and interaction terms. $\psi_L (\psi_R)$ is the spinor for
left (right) handed fermions. $\gamma$ is the Immirzi parameter. $H
= \frac{\dot a}{a}$ is the Hubble rate.

The key ingredient that dynamically cancels the cosmological
constant arises from the four-fermion interaction in the r.h.s of
(\ref{ham}).  This effect arises from an interplay between general
covariance and non-perturbative quantum mechanics.  General
covariance guarantees the four-fermion interaction.  What about the
non-perturbative quantum mechanics?  We see that the effective
coupling of the four-fermion interaction becomes large for small
values of the scale factor (ie. at early times).  The form of this
Hamiltonian maps directly into the BCS Hamiltonian of
superconductivity, except it is the gravitational field that is
playing the role of the phonons. As a result, just like in the BCS
theory (see eg. \cite{Fetter_Walecka}), an energy gap $\Delta$ opens
up which reflects the instability of the ground state associated
with the bare cosmological constant.  An effective cosmological
constant with a lower energy is generated from the formation of the
gap.  To obtain the gap, we diagonalize the fermionic part of this
Hamiltonian by expanding the fermions in normal modes and using a
Boguliubov transformation. The resulting Hamiltonian is:

\begin{eqnarray}
\mathcal{H} &=& -\frac{3}{\kappa}H^2 + \frac{1}{\kappa}(\Lambda_0 -
\Lambda_{corr})
        + \nonumber \\
         && \int \frac{d^3k}{(2\pi)^3} \sqrt{E_k^2 + \Delta^2}(m_k + \bar m_k + n_{-k} + \bar
        n_{-k})
\end{eqnarray}

where the non-perturbative correction to the bare cosmological
constant is\footnote{$\Delta$ is obtained by solving the gap
equation obtained in \cite{alexander-2006} in a self-consistent
manner}:

\begin{eqnarray}\label{gap1}
    \Lambda_{corr} & = & 2 \Delta^2 \nonumber \\
    \Delta & = & \frac{2\hbar\omega_D \exp^{\frac{\nu}{2}}}{\exp^{\nu} - 1}, 
    \qquad \left(\nu = \frac{2}{\kappa\, a^3 k_f^2}\frac{\gamma^2 + 1}{\gamma^2} \right)
\end{eqnarray}

$k_f$ is the fermi energy, $E_k$ is the energy of the $k^{th}$ mode
of the condensate and $\gamma$ is the Immirzi parameter. $m_k (n_k)$
and $\bar m_k (\bar n_k)$ are the creation (annihilation) operators
for the condensate of the left and right-handed fermions
respectively. The $a^3$ factor in $\Delta$ comes from the fact that
the density of states in an expanding universe scales as the
3-volume. We see that the last term in the Hamiltonian constraint
corresponds to the quantized expression for a scalar field
condensate, $\phi_{c}$, with mass $\Delta$ \footnote{To be precise
we note that there are \emph{two} scalar fields, corresponding to
the two pairs of annihilation and creation operators. However in the
following we use only one scalar field for simplicity. Noting that
the left handed massless fermions are the antiparticles of the right
handed ones, we can conjecture that this expression is the quantized
form of a \emph{complex} scalar field, which would imply that we are
dealing with an axion}. Replacing this with the classical expression
for a scalar field we get:

\begin{equation}
    \mathcal{H} = -\frac{3}{\kappa}H^2 + \frac{1}{\kappa}(\Lambda_0 - 2\Delta(a)^2)
        + \frac{1}{2}\dot\phi_c^2 + \frac{1}{2}\Delta(a)^2 \phi_c^2 = 0
\end{equation}

It is important to keep in mind that $\phi_c$ is not a fundamental
scalar field. Its annihilation and creation operators ($m_k$ and
$n_k$) correspond to excitations of the condensate. This leads to
the first Friedmann equation with a time-dependent correction to the
cosmological constant and a scalar field as our matter:

\begin{equation}\label{friedmann_eom}
    3\left(\frac{\dot a}{a}\right)^2 = \Lambda_0 - 2\Delta(a)^2
        + \frac{1}{2}\dot\phi_c^2 + \frac{1}{2}\Delta(a)^2 \phi_c^2
\end{equation}

after setting $\kappa = 1$.

The equation of motion for a scalar in a FRW background is:

\begin{equation}\label{scalar_eom}
    \ddot \phi_c + 3\frac{\dot a}{a}\dot \phi_c + \Delta(a) \phi_c^2 = 0
\end{equation}

We see that the energy gap (\ref{gap1}) increases monotonically with
$a$. From this we can guess the qualitative behavior of the scale
factor.  As long as the initial value of the scale factor is such
that $2 \Delta^2 < \Lambda_0$, then from (\ref{friedmann_eom}) we
see that the right hand side will be positive definite resulting in
an inflating universe. The Hubble rate plays the role of friction
for the scalar field. As time develops the friction will drive $\dot
\phi_c$ to reach zero.  From then until inflation ends, $\phi_c$
will be a constant. Eventually the scale factor becomes large enough
and the right side of (\ref{friedmann_eom}) will start to decrease.
$H$ will then decrease and reach its minimum when:

\begin{equation}\label{critical_gap}
 \Lambda_0 = 2 \Delta^2  -  \frac{1}{2}\Delta^{2}\phi_{c}^{2} -
 \frac{1}{2}\dot \phi_c^2
\end{equation}

$\phi_c$ will then start rolling down the potential hill again,
which is becoming steeper because $a(t)$ and hence $\Delta$ is still
increasing. This presence of the scalar condensate coupling in the
r.h.s of (\ref{critical_gap}) means that when the system dynamically
relaxes to $\Lambda_{eff} = 0$,   it is in a state in which the
energy density of the effective cosmological constant
$\Lambda_{eff}=\Lambda_{0}-2 \Delta^2 $ traces the energy density of
matter.   This condition is similar to the relaxation mechanism due
to backreaction of IR gravitational waves in which the backreaction
effects ceases to negate the cosmological constant and one reaches a
scaling solution where the energy density of matter and radiation
traces the effective cosmological constant\cite{brandenberger-2002,
abramo-1999}. We will see in the next section that once the
cosmological constant is canceled the tracking solution is
dynamically reached without any fine tuning and the cosmological
constant will remain vanishingly small.

We can see that the kinetic energy of the scalar field will
dissipate eventually, due to a small but non-zero $H$. $H = 0$ is
the late-time attractor for this system. As $a(t)$ increases, the
R.H.S. of (\ref{friedmann_eom})will decrease and eventually reach
zero. The solution is stable with respect to perturbations around
this point because of the presence of the gap.

The expression (\ref{critical_gap}) allows us to calculate the value
of the scale factor at the end of inflation. For large $a$, $\Delta
\sim a^3 M_{pl}^2 k_f^2 $. Then from (\ref{critical_gap}) we have:

\begin{equation}\label{final_a}
    a_f = \left ( \frac{\Lambda_0 M_{pl}^2}{2 E_D^2 k_f^4} \right)^\frac{1}{6}
\end{equation}

where $E_D = \hbar \omega_D$. Then for the number of e-foldings we
find:

\begin{equation}\label{efoldings}
    N = ln\left( \frac{a_f}{a_i} \right) \sim - \frac{1}{6} ln(E_D^2 k_f^4)
\end{equation}

where we have set the scale factor at the beginning of inflation
$a_i = 1$. If we assume that $E_D \sim M_{pl}$ and $N = 60$ then
this implies that $k_f \sim e^{-90}$.

\section{Numerical solution and results}

For our numerical calculation we work in Planck units ($\kappa \sim
M_{pl}^2 = 1$). We set $E_D$ and $k_f$ to be $M_{pl} \sim 1$. We
must emphasize that the qualitative behavior is completely
independent of the values of these parameters. In particular, if we
set $k_f = e^{-90}$ we would get 60 e-foldings. It is reasonable to
assume that the bare cosmological constant cannot exceed $M_{pl}^4$
and thus we set $\Lambda_0 = M_{pl}^4 \sim 1$. Then the expression
for the gap becomes:

\begin{equation}\label{gap}
    \Delta = 2\frac{\exp^{\frac{1}{a(t)^3}}}{\exp^{\frac{2}{a(t)^3}} - 1}
\end{equation}

\begin{figure}[htb]
\includegraphics[scale=0.6]{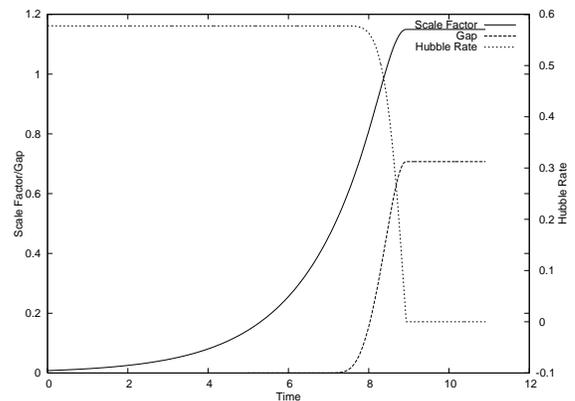}
\caption{\label{fig:gap} Scale factor, hubble rate and condensate
gap as a function of time}
\end{figure}

We solved equations (\ref{scalar_eom}) and (\ref{friedmann_eom})
numerically. An analytic solution is not possible because of
non-analytic form of the gap (\ref{gap}). Fig. \ref{fig:gap} shows
the behavior of the scale factor and the hubble rate as a function
of time.

We find that initially the universe undergoes inflationary expansion
(indicated by the constant value of the hubble rate). When the gap
becomes large enough to cancel out the bare cosmological constant,
inflation ceases.

\begin{figure}[htb]\label{scalar_figure}
\includegraphics[scale=0.6]{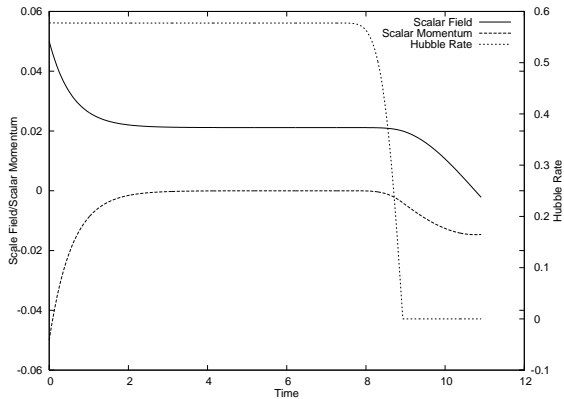}
\caption{\label{fig:scalar} Scalar field and momentum}
\end{figure}

The behavior of the scalar field and momentum is in accord with the
expectations outlined in the previous section. The scalar field
increases or decreases initially depending on the sign of the
initial value of the scalar momentum. It quickly levels off to a
constant value for the rest of the inflationary period, as the
momentum is driven towards zero by a positive $H$ and stays there
until inflation ends. This behavior is independent of the initial
values (which ranged from $0.5$ to $-0.5$ in various runs) and shows
that during inflation the Hubble rate during inflation is always
$\sqrt{\Lambda_0}/3$. In fact, the scalar field plays no role in the
relaxation of the bare cosmological constant. A numerical
calculation setting $\phi_c = \dot \phi_c = 0$ confirms this. Fig.
\ref{fig:scalar} shows the scalar field evolution for one set of
initial values.

\section{Discussion}

In a universe filled with fermions and with a positive cosmological
constants is unstable. Their exists an interaction between fermions
propagated by torsion at the level of the effective field theory.
This interaction leads to the formation of Cooper pairs and a
condensate forms whose free energy is lower than that of the
deSitter background. Consequently the bare cosmological constant,
which we identify to be the free energy of the deSitter background,
is lowered by an amount proportional to the square of the condensate
gap. We have cosmic expansion because initially the gap does not
cancel out the bare cosmological constant completely. The size of
the gap depends on the 3-volume. Hence as the expansion occurs the
effective cosmological "constant" becomes smaller, until eventually
after a period of inflation we emerge from the deSitter vacuum into
flat Minkowski space, where $H \sim 0$.

The number of e-foldings during inflation is given by
(\ref{efoldings}) and can be tuned by adjusting $E_D$ and $k_f$. The
behavior of the scale factor is also independent of the scalar field
evolution.

There are three free parameters in our model. The bare cosmological
constant $\Lambda_0$, the fermi energy $k_f$ and the Debye energy
$E_D$. In a condensate $E_D$ is the cutoff frequency and is
determined by the lattice size. In a cosmological context therefore
we can speculate that it should be $\sim M_{pl}$. $k_f$ can
constrained according to (\ref{efoldings}) to be $\sim e^{-90}$.
$\Lambda_0$ determines the Hubble rate during inflation. From the
WMAP data \cite{spergel-2006}, the upper limit on $H/M_{pl}^2$ is
constrained to be $10^{-4}$. From this we can deduce that the bare
cosmological constant, needs to be fixed by hand to be $\Lambda \sim
H^2 \sim 10^{-8} M_{pl}^4$ in order to conform to observations.

We have presented here a non-perturbative mechanism which relaxes
the bare cosmological constant to zero.  As a bonus we find that the
relaxation is accompanied by an inflationary period. The duration of
inflation is determined by two parameters ($E_D$ and $k_f$) whose
precise determination requires physics beyond the standard model.
The lack of fine-tuning is demonstrated by the fact that the
solution has an attractor with $H=0$ independent of the values of
the free parameters.

The scalar field discussed here is an emergent degree of freedom.
After inflation, oscillations of this field can lead to reheating.
However, to what extent this would be a viable description of the
post-inflationary period remains to be seen in future work.

\begin{acknowledgements}
We would like to thank Robert Brandenberger, Michael Peskin and
Tirtho Biswas for many helpful discussions and suggestions.
\end{acknowledgements}

\bibliographystyle{prsty}

\bibliography{chiral_condensate5}

\end{document}